\documentclass[floats,aps,showpacs,twocolumn]{revtex4}

\usepackage{epsfig}
\def\8{\infty}
\def\oh{\frac{1}{2}}

\def\d{\partial}

\def\undertext#1{\vtop{\hbox{#1}\kern 1pt \hrule}}

\def\VEV#1{\left\langle\,#1\,\right\rangle}

\def\be{\begin{equation}}
\def\ee{\end{equation}}
\def\bea{\begin{eqnarray} & &}
\def\eea{\end{eqnarray}}

\def\rf#1{(\ref{#1})}

\def\rf#1{(\ref{#1})}

\def\rfs#1{Eq.~\rf{#1}}

\begin{document}
\title{A superconductor-insulator transition in a one-dimensional array of Josephson junctions}

\author{V. Gurarie$^{1,2}$ and A.M. Tsvelik$^2$}
\affiliation{$^1$Department of Physics, University of Oxford, 1 Keble Road, Oxford OX1 3NP, UK\\
$^2$Department of Physics, Brookhaven National Laboratory,
  Upton, NY 11973-5000, USA}

\begin{abstract}
We consider a one-dimensional Josephson junction array, in the
regime where the junction charging energy is much greater than the
charging energy of the superconducting islands. In this regime we
critically reexamine the continuum limit description  and
establish the relationship between parameters of the array and the
ones of the resulting sine-Gordon  model.  The later model is
formulated in terms of quasi-charge. We argue that despite
arguments to the contrary in the literature, such quasi-charge
sine-Gordon description remains valid in the vicinity of the phase
transition between  the insulating and the superconducting phases.
We also discuss the effects  of random background charges, which
are always present in experimental realizations of such arrays.

\end{abstract}


\maketitle

\section{Introduction}

The standard description of Josephson junction arrays is given by
the theory by Bradley and Doniach (BD) \cite{BD}. This theory
predicts that a Josephson
junction array can either be in superconducting or insulating
regime, depending on the ratio of the Josephson to the charging
energy of the superconducting grains. The two phases are separated
by the Kosterlitz-Thouless transition.

The theory of Bradley and Doniach assumes that the gate voltage
applied to the arrays is equal to zero. This implies that the
electrostatic energy of the islands is minimized when their charge
is equal to the integer number of the elementary charge of Cooper
pair. More recent work of Glazman and Larkin (GL) \cite{GL}
demonstrates that the gate voltage could be another important
parameter of the Josephson junction arrays, as two additional
phases of the array are possible when it is varied.

%The properties of the array must be periodic in gate voltage.
%Indeed, increasing the voltage by the suitable amount will simply
%increase the number of Cooper pairs on each island by one.

The BD transition was recently observed in one
dimensional Josephson junction arrays \cite{CDH}. However, the
results of Ref.~\cite{CDH} show only qualitative agreement with
the theories of Refs.~\cite{BD} and \cite{CDH}.

First of all,
the BD theory completely neglects the capacitance of the Josephson
junctions $C$ as opposed to the capacitance of the superconducting
islands $C_0$. GL does take nonzero $C$ into account, but only as
a small perturbation. In experiment, however, $C \gg C_0$. Not
only does it mean that the junction capacitance terms are large,
but it also leads to the long ranged Coulomb interactions between
grains which are far away from each other, which is something both
BD and GL neglect. As a result, the point in the parameter space
where the transition happens is experimentally far from the
theoretical point predicted by the BD theory.

It is therefore of some interest to develop a theory of the
Josephson junction arrays which would take the condition $C \gg
C_0$ into account. In fact, such theories were already proposed in
the context of both tunnel \cite{LA} and Josephson \cite{benjacob}
junction arrays. The main idea of this approach is to concentrate
on the dynamics of charge of the islands as opposed to their
superconducting phases. As a starting  point of the theory, one
takes the problem of single Josephson junction with nonzero
capacitance $C$ treated in the full quantum mechanical fashion.
The solution of this problem is a Josephson junction described by
its quasicharge \cite{Likh}. For a system of connected junctions
this description in the continuous limit yields the sine-Gordon
Lagrangian describing dynamics of the quasicharge.

The sine-Gordon model in (1+1)-dimensions is one of the best
studied models of field theory. This model admits a quantum phase
transition: the excitation spectrum depends on the coupling
constant and becomes gapless when this constant exceeds a certain
critical value. In the underlying Josephson array this transition
corresponds to the superconductor-insulator transition
 predicted by BD. We have to note, however,
 that according to the opinion widely circulating in the literature the
 quasicharge sine-Gordon description of
the Josephson junction arrays
loses its validity at large coupling constants and therefore cannot ascertain the
existence of the transition \cite{benjacob}.

In this paper we critically examine the conditions of validity of
the sine-Gordon description of Josephson junction arrays. We
conclude that the quasi-charge sine-Gordon description remains
self-consistent in the region of large couplings. Hence one can
use it to get quantitative information about the array even in the
vicinity of  the transition.

We have to emphasize that the BD theory also describes the
transition in terms of the sine-Gordon equation. However, being
written in the limit $C \ll C_0$, the parameters of that equation
are different from the one extracted from the sine-Gordon equation
considered in this paper. We conclude therefore that while the
universality class of the transition does not depend on whether
$C$ is large or small, the details of the description do depend on
it.

Secondly, contrary to Ref.~\cite{GL}, the experiments failed to
observe any additional phases of the array due to the change in
gate voltage. Most likely, this is the result of the presence of
random background charges in the realistic Josephson junction
arrays, which lead to the voltage being randomized along the
length of the array \cite{Explain}. An important outstanding
question is thus the effect the random charges may have on the
array and how their presence affects the transition.

The problem of Josephson junction arrays with random background
charges is equivalent to the problem of randomly pinned charge
density waves \cite{FL}, which was extensively studied in the
literature. The best understood case is that of classical pinned
charge density waves, which corresponds to the array deeply in the
insulating phase. In that regime, the random background charges
lead to the AC conductance of the array \cite{AR,Fogler,GC}
\begin{equation}
\sigma(\omega) \propto \omega^4, \ \omega \ll \omega_p
\end{equation}
on the condition that the array length is much longer than the
so-called Larkin length $l_p$ \cite{Larkin}. Here $\omega_p$ is
the pinning frequency. Additionally, for a special value of the
sine-Gordon coupling parameter $\beta^2 = 4 \pi$ (see later in the
text for a precise definition of $\beta$), the AC conductance is
known to go as \cite{berezinskii}:
\begin{equation}
\label{ber} \sigma(\omega) \propto \omega^2\ln\omega
\end{equation}
This can in principle be checked experimentally.  In this paper we
discuss a relationship between $\beta, l_p$, $\omega_p$ and
parameters of the junction array.

Unfortunately, at other values of the coupling parameter $\beta$,
the solution to the random background charge problem is not known.
Therefore, we are not able at this time to discuss how the
presence of the random background charges affects the
superconductor-insulator transition.

The rest of this paper is organized as follows. In the next
chapter we derive the quasicharge sine-Gordon equation. Then we
discuss its applicability and show that it can indeed describe the
transition superconductor-insulator transition. In the end we
discuss how the random background charges affect the properties of
the array.

\section{Derivation of the Model}

Let us  consider a one-dimensional array of  weakly coupled Josephson
junctions (see Fig. 1). A single junction is described by the
Hamiltonian
\begin{equation}
H=- {E_c \over 2} {\d^2 \over d \phi^2} + \oh E_j \cos(\phi)
\end{equation}
where $E_c={(2 e)^2 \over C}$ is the charging energy of the
junction, $E_j$ is the Josephson energy and  $\phi$ is the phase
difference on the junction.

\begin{figure}[ht]
\begin{center}
\epsfxsize=0.45\textwidth \epsfbox{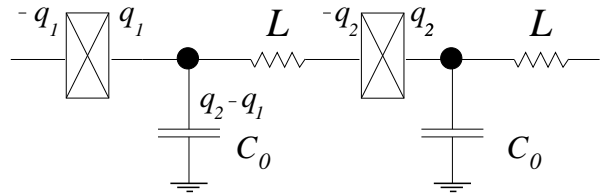}
\end{center}
\caption{An equivalent scheme of an elementary cell of the array.
Black dots denote the superconducting islands.}
\end{figure}

We work in the limit $E_j \gg E_c$. In this regime, the energy
spectrum  consists of narrow  bands separated by gaps (see
Ref.~\cite{LA,Likh}). The interband splitting frequency  is
estimated as
\begin{equation}
W= 2 \sqrt{E_c E_j} \label{Lambda}
\end{equation}
We shall restrict our consideration  to the lowest band. Therefore
$W$ will serve as the ultraviolet cut-off in our effective theory.
The energy eigenvalues in the lowest band are  labelled by the
quasicharge $q$:
\begin{equation}
E= E_b \left[1-\cos(2 \pi q)\right] \label{E}
\end{equation}
Here
\begin{equation}
E_b = 16 E_j \left( {E_c \over  \pi^2 E_j }\right)^{1 \over 4}\exp
\left( -2 \sqrt{E_j \over E_c} \right) .
\end{equation}

The junction array consists of junctions separated by
superconducting islands. Following \cite{benjacob}, we treat the
array in the adiabatic approximation, so that the $n$-th junction
is characterized by a slowly varying function  $q_n(t)$. The
characteristic frequencies are supposed to be much smaller than
the interband splitting $W$. The charge on the $n$-th junction  is
$2 e \times q_n$.  Since the total charge in the unit cell is
zero, the charge on the wire connecting the two junctions is
$2e(q_n - q_{n +1})$. The Coulomb  energy of the wire  is
\begin{equation}
E_{\hbox{Coulomb}} = {(2 e)^2 \over 2 C_0 }\sum_n (q_n - q_{n
+1})^2 \approx \int dx~a {(2 e)^2 \over 2 C_0 } q_x^2 \label{C}
\end{equation}
where $a$ is the size of the island. The inductive energy of the junction is
\begin{equation}
E_{\hbox{Inductive}} =\oh \sum_n (2 e)^2 L \dot{q}_n^2 \label{L},
\end{equation}
where $L$ is the inductance of the islands. The experiments
conducted in Ref.~\cite{hysteresis} indicate that real arrays have
a very considerable inductance ($L \gg {\hbar^2 \over e^2 E_j}$),
though its microscopic origin  remains somewhat obscure.

Combined, Eqs.(\ref{E},\ref{C},\ref{L}) produce the Lagrangian
describing the array,
\begin{equation}
{\cal L} = \int dx \left\{ {(2 e)^2 \over 2 a} L \dot q^2 - a {(2
e)^2 \over 2 C_0} q_x^2 - {E_b \over a}  \left[ 1-\cos(2 \pi q)
\right] \right\}
\end{equation}
By introducing a new variable $Q = 2\pi q/\beta$
 the latter Lagrangian can be rewritten in the form of the sine-Gordon model:
\begin{equation}
\label{SG} {\cal L} = {\hbar v_c^{-1}} \int dx~\left[ {1 \over
2} \dot Q^2 - \oh v_c^2 Q_x^2 - {m_0^2 \over {\beta^2}} \left(1-\cos
\beta Q\right) \right]
\end{equation}
where
\begin{equation}
m_0^2={\pi^2 E_b \over e^2 L },
\end{equation}
\begin{equation}
v_c = {a \over \sqrt{L C_0}} ,
\end{equation} and
\begin{equation}
\label{beta} \beta^2= {(2 \pi)^2 \hbar v_c C_0 \over (2 e)^2 a} =
{(2 \pi)^2 \hbar \over (2 e)^2 } \sqrt{C_0 \over L} = 2 \pi
R_Q\sqrt{C_0 \over L},
\end{equation}
where $R_Q = h/4e^2$ is the quantum of Cooper pair resistance.

The sine-Gordon equation \rf{SG} goes through a quantum phase
transition as $\beta^2$ is tuned through $8 \pi$. Let us estimate
whether this value of $\beta$ can realistically be achieved.
Assuming that the island has a form of a thin superconducting wire
we can use for its impedance $\sqrt{L/C_0}$ the formula derived by
Likharev \cite{old}:
\begin{equation}
\sqrt{L/C_0} = \frac{4\pi}{c}\left[1 + \frac{2\pi\lambda^2}{A\ln(D/d)}\right]^{1/2}
\end{equation}
where $\lambda$ is the penetration depth, $A$ is the cross-section
area of the wire, $d$ is its thickness and $D$ is the distance
from the gate. Taking into account that $\hbar c/e^2 \approx 137$
we get
\begin{equation}
\frac{\beta^2}{8\pi} \approx \frac{4.3}{\left[1 + \frac{2\pi\lambda^2}{A\ln(D/d)}\right]^{1/2}}
\end{equation}
We see that $\beta^2$ can be driven through the transition by changing $\lambda$ which can be achieved by adding nonmagnetic impurities \cite{private}.

\section{Applicability limits of the sine-Gordon model}

As we have mentioned above,  the sine-Gordon model Eq.~\rf{SG} is
one of the best studied models of one-dimensional field theory.
Its spectrum and behavior of correlation functions dramatically
depend on the value of coupling constant $\beta^2$, because
$\beta^2$ controls the renormalization group (RG) dimension of the
cosine term in Eq.~\rf{SG}. At $\beta^2 < 8\pi$, the cosine term
is relevant,  all excitations have spectral gaps and the system is
an insulator. At $\beta^2 < 4\pi$ the spectrum consists of
electrically charged solitons and antisolitons and their  neutral  bound
states, while at $8 \pi > \beta^2 > 4\pi$ the bound states disappear.
At $\beta^2
> 8\pi$, the cosine term is irrelevant and the spectrum becomes gapless. This is a superconducting regime; in the absence of disorder it allows
ballistic transport through the system.

 Let us critically examine conditions of validity of theory \rf{SG}. It is based on the following assumptions:
\begin{itemize}
\item
The harmonicity of the effective potential. The potential
 acquires cosine form \rf{E} in the
limit $E_J \gg E_c$. The presence of higher harmonics will affect
the integrability of the sine-Gordon model, but will not change
the fact of the transition since these harmonics will become
irrelevant even sooner than the primary one. \item The
abiabaticity. All characteristic energies of sine-Gordon model (in
particular, the spectral gaps) must be much smaller than the
interband splitting $W$ of Eq.~\rf{Lambda}. \item The continuous
approximation. The discrete array is replaced by the continuous
one. This approximation requires that characteristic wave vectors
are much smaller than $1/a$ and the energy gaps are much smaller
than $\Lambda=\hbar v_c/a$. \item Absence of dissipation. It is
assumed that islands are superconducting and there are no normal
resistors in the scheme. \item We assume that all characteristic
frequencies are much smaller than the plasma frequency in the
islands so that electric charge spreads instantaneously through
the entire island.
\end{itemize}

Let us look closer at the requirement of adiabaticity and validity
of the continuous approximation. The spectrum of the sine-Gordon
model consists of particle branches with the relativistic
dispertion
\begin{equation}
\omega^2=v_c^2 k^2 + m_j^2/\hbar^2,
\end{equation}
where the spectrum consists of solitons $s$ and antisolitons $\bar
s$ with $m_s = m_{\bar s} = M$ and (at $\beta^2 < 4\pi$) their
bound states (breathers) with spectral gaps
\begin{equation}
m_j = 2M\sin(\gamma j/2), ~~ j = 1, \dots,  [\pi/\gamma], ~~
\gamma = \frac{\pi\beta^2}{8 \pi - \beta^2}
\end{equation}
The largest spectral gap is of order of $M$. The self-consistency
of the theory requires that its value must not exceed the cut-off.
According to \cite{zam}, we have
\begin{eqnarray}
{M \over \Lambda}  = \frac{2\Gamma(\gamma/2)}{\sqrt{\pi}\Gamma(1/2
+ \gamma/2)}\left[\frac{m_0^2\Gamma(1 -
\beta^2/8\pi)}{16\Lambda^2\Gamma(1 + \beta^2/8\pi)}\right]^{1
\over 2 - \beta^2/4\pi}
\end{eqnarray}
The following inequalities must be fulfilled:
\begin{equation}
M \ll \Lambda \ll W
\end{equation}

Inequality $M \ll \Lambda$ is equivalent either to
\begin{equation}
m_0 \ll \gamma\Lambda, ~~ E_b \ll 1/e^2L
\end{equation}
at small $\beta^2$ or to
\begin{equation}
m_0 \ll \Lambda, ~~ E_b \ll e^2/C_0
\end{equation}
at $\beta^2 \approx 8\pi$. The validity of the continuous
approximation requires that $\Lambda \ll W$, that is
\begin{equation}
{\hbar^2 \over C_0 L} \ll  E_j E_c, ~\hbox {or}~ \beta^2 \ll W
C_0/4e^2
\end{equation}
From the previous inequalities it follows that the only condition
on the latter quantity is $WC_0/e^2 \ll \left( {E_j \over E_c}
\right)^{1 \over 4} \exp[2 \sqrt{ E_j/E_c}]$ which still leaves
room for $\beta^2 \approx 8\pi$.

\section{Phase Transition}

As we have mentioned above, one way to drive the system through
the transition is by changing $\beta^2$. In fact, it is also
possible to change $E_b$ or equivalently $m_0$, to go through the
transition. This  would correspond to the experimental set up
\cite{CDH}, where  $E_j$ is lowered by applying the
external magnetic field to the junctions in the SQUID geometry,
leading to the increase of $E_b$ and $m_0$.

The Renormalization Group (RG) flow diagram of the sine-Gordon equation is shown on Fig.
2. If $\beta^2 < 8 \pi$, the sine-Gordon equation is always in the
massive regime. This corresponds to the insulating behavior of the
array. However if $\beta^2 > 8 \pi$, two regimes are possible. For
sufficiently large $m_0$, the behavior is still massive. For $m_0$
smaller than a certain critical value $m_c \propto \Lambda \left(
\beta^2-8 \pi \right)$, the cosine term in \rf{SG} becomes
irrelevant and can be neglected. The sine-Gordon equation becomes
massless. This corresponds to the part of the diagram whose RG
flow lines end on the $m_0=0$ axis. In this regime the charge
density wave propagates ballistically.

\begin{figure}[ht]
\begin{center}
\epsfxsize=0.45\textwidth \epsfbox{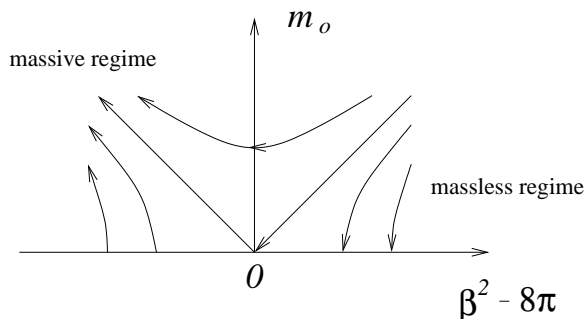}
\end{center}
\caption{The RG flow diagram of the sine-Gordon equation. One can
tune through the transition by either changing the value of
$\beta$ or $m_0$. }
\end{figure}

One of the more experimentally relevant  predictions of the theory for the
ballistic  regime  follows from neglecting the cosine term in
Eq.~\rf{SG}. Then the Josephson junction array can be described by
the following very simple Lagrangian
\begin{equation}
{\cal L}  = {4 \pi^2 \over \beta^2} {\hbar v_c^{-1}} \int
dx~\left[ {1 \over 2} \dot q^2 - \oh v_c^2 q_x^2 \right]
\end{equation}
This is none other than the so-called Luttinger liquid with the
Luttinger parameter $g={\beta^2 \over 4 \pi}$. It is well known
 \cite{FK} that its conductance is given by
\begin{equation}
G={ (2 e)^2 \over h} g = {\beta^2 \over 4 \pi} {(2 e)^2 \over h}
\end{equation}
As $\beta$ is lowered (or $m_0$ is increased, according to Fig 2),
the conductance is decreased until it reaches its critical value
(at $\beta^2=8 \pi$),
\begin{equation}
G^*= 2 {(2 e)^2 \over h}={2 \over R_Q}.
\end{equation}
If $\beta$ is decreased further, the array enters the regime of
the Coulomb blockade and becomes an insulator.

\section{Random Background Charges}

In the presence of random background charges, the charging energy
of the superconducting islands \rfs{C} gets modified,
\begin{equation}
E_{\hbox{Coulomb}} = \int dx~a {(2 e)^2 \over 2 C_0 } \left( q_x -
V(x) \right)^2 \label{C1},
\end{equation}
where $V(x)$ is a time independent random function of the
coordinate $x$ with short ranged correlations
\begin{equation}
\VEV{V(x) V(y)} = V_0 \delta(x-x').
\end{equation}
It is convenient to shift the variables $q(x) \rightarrow
q(x)+\int_0^x dy~V(y)$ to find the following Lagrangian for the
Josephson junction array,
\begin{equation}
\label{SG1} {\cal L} = {\hbar \over 2} \int dx~\left[ {1 \over
v_c} \dot Q^2 -  v_c Q_x^2 - {2 m_0^2 \over {\beta^2}}
\left\{1-\cos \left( \beta Q + \chi \right) \right\} \right],
\end{equation}
where
\begin{equation}
\chi(x) = 2 \pi \int_0^x dy~V(y).
\end{equation}
It is clear that $\chi(x)$ is a Brownian motion;
$\sqrt{\VEV{\chi^2(x)}}= 2 \pi \sqrt{x V_0}$. At distances of the
order of $l \propto V_0^{-1}$ the phases of the cosine term in
\rfs{SG1} become completely uncorrelated, so at length scale much
bigger than that we can consider $\cos\left\{\chi(x)\right\}$ to
be white noise in space. $l$ is the length at which the random
background charge accumulates to be of the order of the Cooper
pair charge.

Under these conditions, the problem defined in \rfs{SG1} becomes
equivalent to the pinned charge density wave problem studied in a
number of publications \cite{Larkin}. A particularly well understood regime is
that of a Coulomb blockade, $\beta \ll 8 \pi$. There the
sine-Gordon Lagrangian \rfs{SG1} becomes purely classical,
corresponding to the sine-Gordon equation
\begin{equation}
\label{Q0} v_c^{-1} \d^2_t Q - v_c \d^2_x Q - { m_0^2 \over \beta}
\sin \left\{ Q+\chi \right\} =0.
\end{equation}
If $Q_0(x)$ is a time independent solution to \rfs{Q0}, then the
small oscillations around that solution are described by
\begin{equation}
\label{osc25} v_c \d^2_x \delta Q +  m_0^2  \cos \left\{ Q_0+\chi
\right\} \delta Q = v_c^{-1} \omega^2 \delta Q,
\end{equation}
where $\delta Q$ is the amplitude of the oscillations and $\omega$
is their frequency.

It is well known in the literature \cite{AR,Fogler,GC} that this
problem possesses a fundamental frequency called the pinning
frequency $\omega_p$. At $\omega \gg \omega_p$, it is possible to
neglect the cosine term in \rfs{osc25} to find that the
oscillations $\delta Q$ are the plain waves with the wave vector
$k$ and with the frequencies $\omega \propto v_c k$. At $\omega
\ll \omega_p$, the oscillations are localized in space, and their
spacial extent (localization length) $l_p$ is called the Larkin
(or pinning) length. The pinning frequency is known to scale with
$m_0$ as
\begin{equation}
\label{pin} \omega_p \propto m_0^{4 \over 3},
\end{equation}
while the Larkin length scales as
\begin{equation}
l_p = {1 \over \omega_p} \propto m_0^{-{4 \over 3}}.
\end{equation}
On the condition that the Larkin length is much smaller than the
total length of the array, the conductivity of the array at a
frequency $\omega$ much smaller than the pinning frequency is
given by
\begin{equation}
\label{accond} \sigma(\omega) \propto \rho(\omega), \ \rho(\omega)
\propto \omega^4, \ \omega \ll \omega_p.
\end{equation}
Here $\rho(\omega) \propto \omega^4$ is the probability that there
exists a solution to \rfs{osc25} at a frequency $\omega$ (or the
density of states of Eq.~\rf{osc25}). The formula \rfs{accond} as
well as \rfs{pin} could be checked experimentally.

Finally, at a special value of $\beta^2 =4 \pi$, the \rfs{SG1} can
be solved using a completely different technique
\cite{berezinskii}, to lead to the AC conductivity \rfs{ber}.

At values of $\beta$ other than $\beta^2 \ll 8 \pi$ or $\beta^2 =
4 \pi$, the problem defined in \rfs{SG1} remains unsolved.

\section{Acknowledgements}

 We are grateful to D. Haviland, K. K. Likharev, L. Glazman  and F. H. L. Essler for productive discussions.
 VG acknowledges support from Institute for Strongly Correlated and Complex Systems at BNL  and from EPSRC;
 AMT is supported by  US DOE under contract No. DE-AC02-98CH10886.

\end{document}